\documentclass[10pt,times,mathptm,aps,article,twocolumn,showpacs,showkeys]{revtex4}

\usepackage[]{graphicx}
\usepackage{amssymb}
\usepackage{bm}

\begin{document}

\setcounter{page}{1}

{\it Published in} Journal of the Korean Physical Society

Vol. 47, $N^o$ 91, 30 August 2005, pp. S175-S181

\title[]{\vspace{15mm} Nanostructuring Optical Waveguides by Focused Ion Beam Milling.
\\
Near-Field Characterization.\vspace{5mm}}
\author{Fr\' ed\' eric \surname{Lacour}}
\author{Andrei \surname{Sabac}}
\author{Michel \surname{Spajer}}
\email{michel.spajer@univ-fcomte.fr}
\thanks{Website:\url{http://www.femto-st.fr/fr/Departements-de-recherche/OPTIQUE/}}

\affiliation{\vspace{2mm} Institut FEMTO-ST, D\' epartement d'Optique P. M. Duffieux,
Universit\' e de Franche-Comt\' e, UMR 6174 CNRS, 25030 Besan\c con cedex, France
\vspace{2mm}}

\date[]{Received 6 September 2004}

\begin{abstract}
{\it Available online at }\url{http://www.kps.or.kr/home/kor/journal/library/search.asp?} 

\vspace{5mm} Nanostructures have become an attractive subject due to many applications,
particularly the photonic bandgap effect observed in photonic crystals. Nevertheless,
the fabrication of such structures remains a challenge because of accurate requirement
concerning  regularity, shape, hole depth etc. of the structure. E-beam lithography
permits a good control of dimensional parameters but needs a 1-step fabrication process.
In our work, we have to combine traditional strip-load waveguides ($SiO_2/SiON/SiO_2$ on
$Si$) and nanostructures whose dimension are totally different. This imposes a 2-step
process where waveguides and nanostructures are successively fabricated. We have at our
disposal different ways to characterize these  nanostructures. A direct aspect control
during and after FIB treatment can be achieved by FIB and SEM imaging. Scanning
near-field optical microscopy (SNOM) is currently the most effective way to test guiding
confinement in such surface structures by detecting the evanescent field.

\end{abstract}



\maketitle

\section{Introduction}

Within the last two decades, the development of technology has permitted structure
fabrication at a submicrometer scale. These nanostructures allow a new control of light,
mainly after the significant progress made in understanding of photonic band gap (PBG)
since the 1980s years \cite{joannopoulos95}. Since then periodic nanostructures (with
dimensions a hundred times smaller than the original waveguides) seem the best way to
develop novel integrated optical devices which dimensions are a hundred times smaller
than the original waveguides. Hunting the losses in such devices is a real challenge
mainly for telecom application. Devices like waveguides or microcavities in photonic
crystals (PhCs) have already been presented and demonstrated \cite{chow00}.
Nevertheless, before all-PhCs components can be developed and replace presently used
components, PhCs and classical optical devices will coexist for a certain time. In this
perspective we were interested in fabricating and characterizing mixed devices, which
combine both photonic structures or nanostructures and classical waveguides.

The most common used PhCs are planar PhCs (lattice of holes in dielectric material or
dielectric pillars in air), which are easier to fabricate than 3D-PhCs which theorically
ensure the best control of light. 2D-PhCs have photonic band gap (PBG) in two
directions, so light has to be confined in the third direction by a multilayer
structure. Therefore high-index materials ($AsGa, AlGaAs$, $Si$ etc.) have been used
because of their large index difference with air which produces large photonic bandgap.
Low-index materials like $SiO_2$ ($n = 1.47$  at $ \lambda = 900 nm$) which are the most
common in integrated optics are generally not studied for PBG application, except for
photonic fibers.

In this work, we fabricate and characterize photonic structures combined with typical
waveguides in materials often used in integrated optics and MOEMS components
\cite{jozwik04}, such as multilayer waveguides $SiO_2/SiON/SiO_2$ on $Si$.

The first part is a theoretical study of the structures based on FDTD (Finite Difference
Time Domain) calculation. These simulations demonstrate the feasibility of a PBG with
low-index materials in specific conditions. Afterwards, we describe the fabrication of
the samples in two steps~: fabrication of the waveguides and of the photonic structures.
We'll show in this part a novel way to fabricate PhCs using a Focused Ion Beam (FIB).
Finally characterization by SNOM probing will be presented.

\section{THEORETICAL STUDY}
\label{sect:theo}

\begin{figure}[!htbp]
        \includegraphics[width=8.5cm]{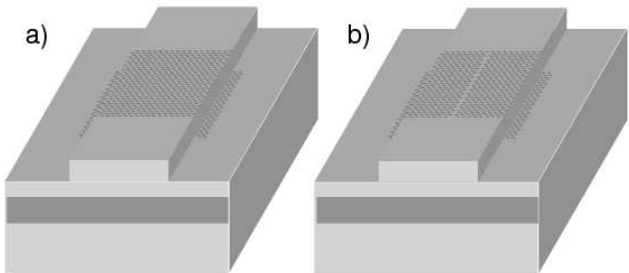}
        \caption{Two studied nanostructures a) without defects b) with a line defect.}
        \label{fig.1}
\end{figure}

Numerical studies are carried out using three commercial software from RSoft. In a first
step a study of infinite PhCs is performed by a PWE (Plane Wave Expansion) software
(RSoft $BandSolve$) to find the PBG and to optimize the parameters of the nanostructure
(hole size and period). In a second step electromagnetic field propagation in the
nanostructures is numerically mapped by a FDTD calculation (RSoft $FullWave$). Rigorous
analysis of geometrical parameters like hole depth or surface roughness requires 3D
calculation, which demands too many computer ressource. Therefore, we shall limit our
study to 2D calculation applied to a 2D equivalent structure, to describe the general
behavior of the 3D structure. The basic structure studied is a triangular lattice of
holes centered on the waveguide as shown in  Fig.\ref{fig.1}a. The equivalent 2D
structure we chose is a vertical guiding slab with a lattice of infinite vertical holes
centered inside.

The free waveguide properties (modes, propagation, losses...) are obtained by a BPM
(Beam Propagation Method) software (RSoft $BeamProp$).

\begin{figure}[!htbp]
        \includegraphics[width=8.5cm]{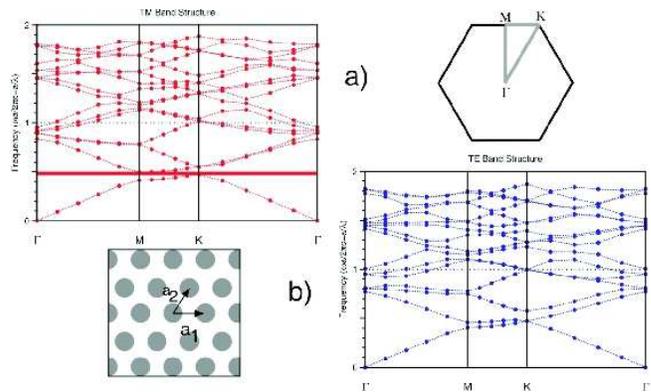}
        \caption{Bandgap diagrams of a triangular hole lattice ($n=1.53$)}
        \label{fig.2}
\end{figure}

\subsection{Photonic bandgaps calculation}

For the wavelength $\lambda=900nm$, effective index of the multilayer waveguide is
calculated by the BPM software to be $n_{eff}=1.489$. Our calculation gives complete
photonic band gap
 in the case of triangular lattice for only one polarization
(TM) (Fig.\ref{fig.2}), for a  hole diameter $d$ between $0.65a$, and $0.82a$ ($a$ is
the period of the lattice). Figure \ref{fig.2} shows the band diagram for $d=0,7a$. A TM
PBG can be obtained for frequencies $0.473<(\frac{\omega .a}{2 \pi
c}=\frac{a}{\lambda})<0.496$. The fact that the PBG in TM mode is very narrow and that
there is no complete band gap in the other polarisation can be explained by the low
index difference between the material and the holes ($\Delta n=n_{eff}-n_{air}=0.489$).

\begin{figure}[!htbp]
        \includegraphics[width=8.5cm]{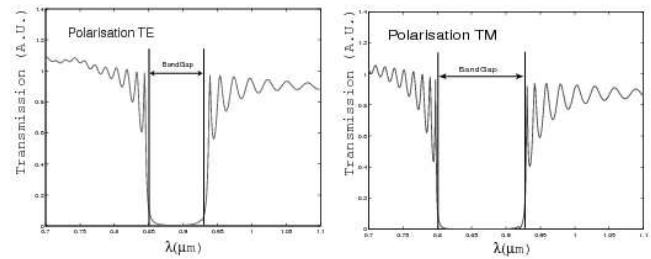}
        \caption{Transmission spectrum of $40 \times 40$ lattice of holes for TE and TM
        polarisation.}
        \label{fig.3}
\end{figure}

Afterwards we only consider  holes of period  $a=360nm$ and diameter  $d=200nm$. These
values were chosen to fit with technological limitations (the ratio $d/a$ can hardly
exceed 0.7 in the milling process) and to obtain a bandgap in the wavelength range
(around 700-980nm) given by the tunable laser (See \ref{sect:char}).

\subsection{FDTD Simulations}

After determining the  parameters of the hole lattice the combination of the
nanostructures with the waveguide can be characterized by FDTD calculation.

Two structures were modelled: the first one is a complete lattice, in the second one a
line of holes has been removed at the center of the waveguide ( Fig.\ref{fig.1}b). It is
expected that the lattice of holes without defects will behave like a mirror for
wavelengths included in the photonic band gap.

\subsubsection{Complete triangular lattice of holes}

Light is injected in the equivalent planar waveguide wherein the hole lattice has been
pierced according to the  previous parameters. In a first step, transmission spectra
have been calculated by injecting a pulse excitation in the direction $\Gamma M$ . Two
samples of this spectrum are presented in Fig.\ref{fig.3}. Although PBG calculation
exhibits no complete band gap for TE polarisation and a narrow one for TM polarisation,
the spectrum shows a band gap for both polarisations. For TE polarisation a fall of
transmission efficiency can be noticed for wavelengths between $848nm$ et $934nm$. In TM
polarisation a similar PBG exists for wavelengths between $800nm$ and $927nm$. It must
be noticed that a partial band gap can exist for a given incidence, even if the crystal
has no complete band gap (independent of the incidence). In the present case, the
spectra of Fig.~\ref{fig.3} correspond to the partial band gaps observed on the vertical
line $M$ of Fig.~\ref{fig.2}. It can be verified that the width given by FDTD is
slightly smaller ($\times 0.8$) than the previous one as shown in table~I. This
discrepancy can probably be explained by the plane wave expansion of the guided mode :
the direction of incidence is not limited to the waveguide axis .

\begin{table}
\caption{Comparaison of obtained band gap between FDTD calculation (FullWave) and PWE
method (BandSolve).}
\begin{ruledtabular}
\begin{tabular}{ccc}
  & BandSolve & FullWave\\
 TE $\Delta(a/\lambda)=$ & 0.0495 & 0.0391\\
 TM $\Delta(a/\lambda)=$ & 0.0795 & 0.0621\\
 TM complete PBG & $\Delta(a/\lambda)=0.023$
\end{tabular}
\end{ruledtabular}
\end{table}

\begin{figure}[!htbp]
        \includegraphics[width=8.5cm]{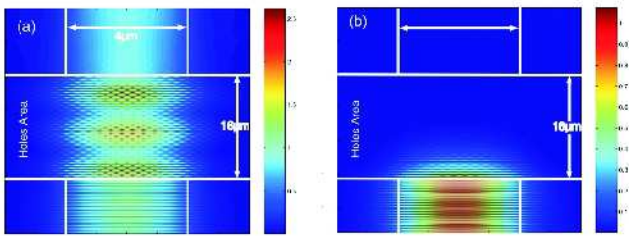}
        \caption{Cartographies of $Ey$ field for TE polarisation
        at $\lambda=800nm$ (a) (outside PBG) and $\lambda=900nm$ (b)
         (inside PBG) for $40 \times 40$ holes triangular lattice
         ($a=360nm$, $d=200nm$).}
        \label{fig.4}
\end{figure}

The fall of transmission increases with the number of periods in the propagation
direction. We found that a minimum of ten periods is necessary to obtain a $0.1 \%$
transmission. For this example, 40 periods have been used.

Then, by launching continuous excitation and waiting for field stabilization in the
structure, a mapping of each field ($Ey$, $Hx$ and $Hz$ for TE polarisation, $Ex$, $Ez$
and $Hy$ for TM polarisation) can be obtained as shown in Fig.\ref{fig.4}. For
$\lambda=800nm$, light can pass through the lattice, but for $\lambda=900nm$, light
seems to be completely reflected.

\subsubsection{Triangular lattice with a line defect}

\begin{figure}[!htbp]
        \includegraphics[width=8.5cm]{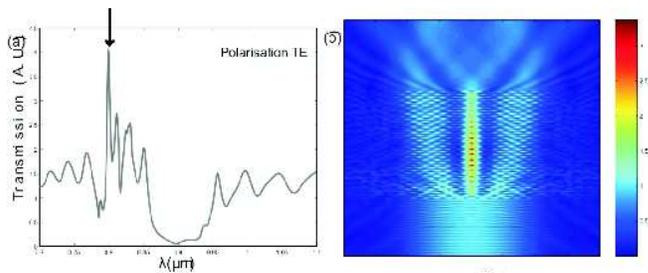}
        \caption{(a) Transmission spectrum of $40 \times 40$ lattice of holes
        with a line defect for TE polarisation. (b) $E_y$ field cartography (2D FDTD
        calculation) for $\lambda =800nm$}
        \label{fig.5}
\end{figure}

We carried out the same calculation for a triangular lattice of holes where one line is
missing. Light propagation is always in the direction $\Gamma M$. Transmission spectrum
shows that the band gap has been perturbed and that some peaks appear (See
Fig.\ref{fig.5}(a)). For a better coupling efficiency between optical waveguide and PBG
waveguide of different width,
 a taper would be desirable (optical waveguide is $4 \mu m$ wide
and a line missing line is wide of one period of the lattice ($360nm$)). Nevertheless 2D
FDTD calculation at the wavelength corresponding to the higher peak ($\lambda =800nm$)
exhibits a high confinement of light in the line defect (See Fig.\ref{fig.5} (b)). A
calculation made at PBG wavelengths shows no light propagation in this line defect: the
result is similar to Fig.\ref{fig.4} (b).

\section{NANOSTRUCTURING OPTICAL WAVEGUIDES}
\label{sect:fab}

Photonic crystal fabrication remains a challenge mainly due to the size and the depth of
engraved holes. In our case, hole diameter
 is $200 nm$ and the depth has to be more than $3 \mu m$ to
overlap the optical guided mode. Moreover, as optical waveguides have been fabricated
beforehand, a good precision is needed to align correctly waveguides and
nanostructures.

In a previous work \cite{lacour04}, we have demonstrated the possibility of
nanostructuring lithium niobate substrate by Focused Ion Beam (FIB) the dimension of
which is less than $50 nm$. Other laboratories have obtained FIB patterns at
submicrometer scale \cite{gierak01} with good optical properties \cite{Dale00,
bostan03}. Furthermore the main advantage of FIB is the ability to drill holes directly
from the sample surface and the direct control of the structures after milling.

\subsection{Waveguide fabrication}

Optical waveguides are an important part in the fabrication of the sample since they
guide light to nanostructures and also provide optical confinement in the third
dimension.

\begin{figure}[!htbp]
        \includegraphics[width=7.0cm]{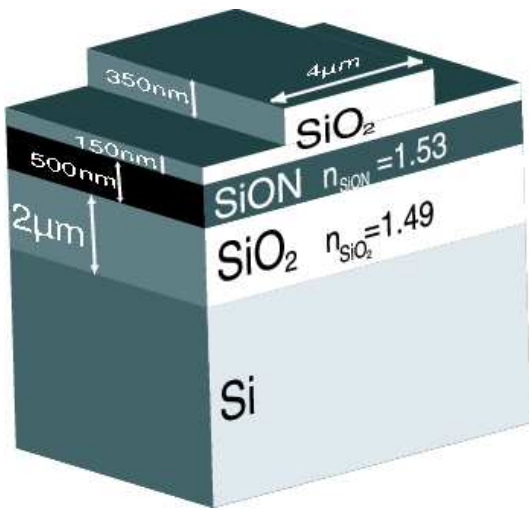}
        \caption{Description of the optical waveguides.}
        \label{fig.6}
\end{figure}

The optical waveguides (Fig.\ref{fig.6}) we have chosen are multilayer waveguides
($SiO_2/SiON/SiO_2$) deposited on a $Si$ substrate. The three layers ($SiO_2$,
thickness: $e_1=3 \mu m$, $SiON$: $e_2=0.5 \mu m$ and $SiO_2$ again: $e_3=0.5 \mu m$)
are deposited by PECVD process. This process allows a good control  over the thickness
and over the optical index of the different layers. After a lithographic masking the
rib waveguides are etched by RIE process on the top $SiO_2$ layer.

For the chosen parameters, BPM calculation have shown that waveguides are single mode
over the wavelength scale which will be used for near-field characterization
($700-950nm$).

\begin{figure}[!htbp]
        \includegraphics[width=8.5cm]{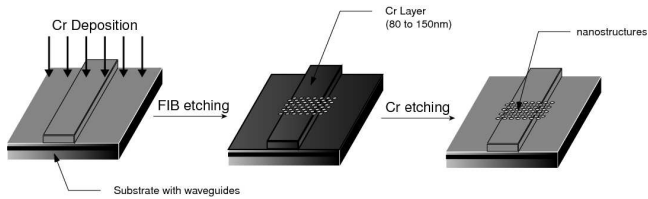}
        \caption{FIB nanostructuring of the waveguides.}
        \label{fig.7}
\end{figure}

\subsection{Nanostructuring waveguides with FIB}
\label{sect:nanostructuring}

The fabrication process (See Fig.\ref{fig.7}) is based on a direct etching of the sample
substrate by FIB bombardement. The samples, on which multilayer waveguides have been
previously fabricated, are chromium coated ($<150 nm$ layer) by electron gun evaporation
to avoid charging,  and grounded with a conductive paste before introduction in the FIB
vacuum chamber ($<10^{-8}torr$). Then the sample is milled using a FIB column
(Orsay-Physics - LEO FIB4400). $Ga^+$ ion are emitted with a current of $2 \mu A$ and
accelerated by a voltage  of $30kV$. The probe current of the focused beam is $66pA$ or
$115pA$ . The Gaussian-shaped spot size is estimated to be $50 nm$ in the best
conditions. The ion beam scans the sample, the deflection field being controlled by a
software to produce the desired pattern.

\begin{figure}[!htbp]
        \includegraphics[width=8.5cm]{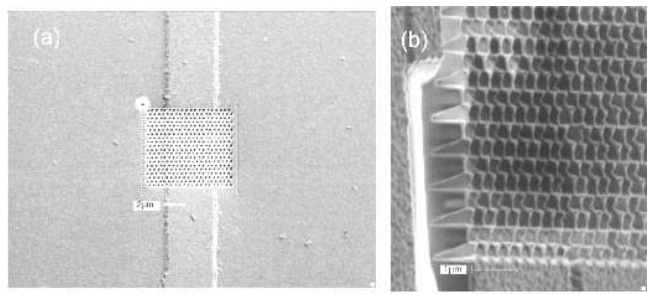}
        \caption{(a) SEM image of a $20 \times 20$ lattice of holes on a waveguide.
        (b) FIB image of a cross section made by FIB milling of the same lattice.
        Measured depth: $e=1.20 \mu m$ (view angle $\alpha =30^o$)}
        \label{fig.8}
\end{figure}

Several triangular lattice of circular holes with previously determined parameters
($a=360nm$ and $d=200nm$) have been milled on the substrate. Fig.\ref{fig.8} shows two
views of a $20 \times 20$ hole lattice fabricated by FIB. The cross section realized by
FIB (Fig.\ref{fig.8} (b)) provides important information about the hole depth (which we
expect to be more than $1 \mu m$ deep), the surface aspect and the hole shape . Hole
shape is not cylindrical as we first wanted but  conical. The upper diameter at the
surface is about $220nm$ and the bottom diameter $120nm$. This kind of problem is a
fundamental drawback of FIB milling due to material redeposition during the scanning
process. More vertical sidewalls can be obtained by using reactive gases during the FIB
process.

\section{Characterization}
\label{sect:char}

We have characterized this fabricated nanostructures using two different methods. A
transmission spectrum of those nanostructures has first been studied by measuring the
ratio between intensities of injected light and recovered light at the waveguide
extremity. The second method is an in-situ characterization of the nanostructures with a
stand-alone near-field microscope working in detection mode. This technique has been
employed by several laboratories, more specifically in the study of high electromagnetic
confinement (See \cite{peeters00, Colas01, Bolzhevolnyi02, Gerard02, Kramper04}).

\subsection{Transmission spectrum}

A tunable Titane-Sapphire Laser is used to scan a wavelength range from $\lambda =700nm$
to $\lambda=980nm$. Its beam is injected into a polarization maintaining optical fiber
which can be oriented to select TE or TM polarization. Light is injected directly in the
optical waveguides from this fiber to limit the optical losses. Injection is controlled
by collecting the guided mode from the end face of the waveguide by a microscope
objective and a CCD camera (Fig.\ref{fig.10}). The transmission spectrum exhibits
variations of about $20\%$ but does not show any clear photonic band gap.

\begin{figure}[!htbp]
        \includegraphics[width=8.5cm]{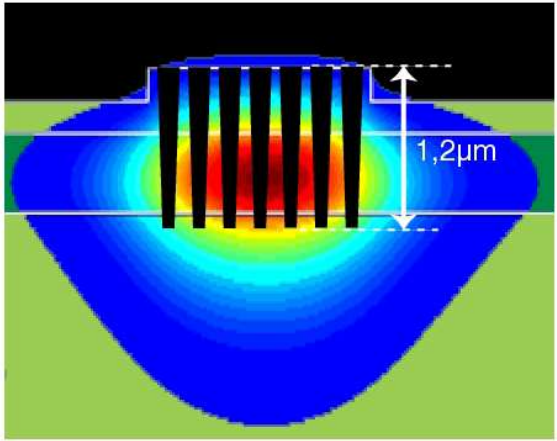}
        \caption{Problems of fabricated structures. Superimposition of BPM calculated
        guided mode and structures.}
        \label{fig.9}
\end{figure}

\begin{figure}[!htbp]
        \includegraphics[width=8.5cm]{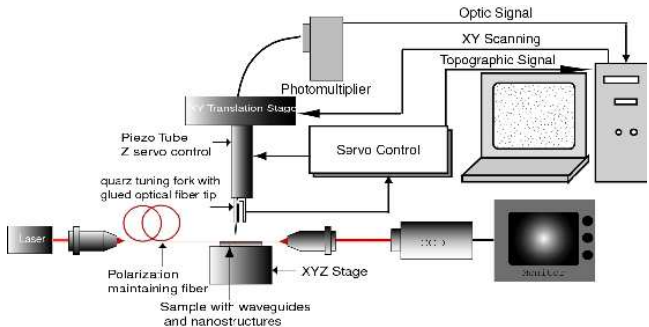}
        \caption{Experimental setup.}
        \label{fig.10}
\end{figure}

The reason seems to be the out-of-plane losses caused by the perforation of the
waveguide which induces a coupling of the guided wave with radiation modes. Recent work
\cite{benisty02, ferrini03, ferrini03b} has demonstrated that out-of-plane losses in
planar PhCs depend on two principal parameters: they occur when the overlap between the
guided mode and the nanostructures is not complete and when the sidewalls of holes are
not vertical.  From the cross-section of Fig.\ref{fig.8} (b) the tilt of the sidewall is
about $2.5^o$ and the holes are deeper than the core layer but do not overlap the mode
completely. Fig.\ref{fig.9} superimposes the hole profile and the BPM calculated optical
mode. Reference \cite{ferrini03} demonstrates that a sidewall tilt bigger than $1^o$ may
have disastrous consequences on the $LiNbO_3$ effect.

Even if a real PBG cannot be observed by transmission spectrum, we'll see that
near-field characterization could show relevant information on these nanostructures.

\subsection{Near-field characterization}

\subsubsection{Experimental setup}

A stand-alone microscope has been used to characterize nanostructures on waveguides
(Fig.\ref{fig.10}). The evanescent wave on top of the structure is collected by an
optical probe made of an optical fiber tapered by the pulling/heating method under
$CO_2$ laser beam. Gluing the probe to a tuning fork ensures the usual distance control
by shear-force feedback. The laser beam is modulated by a chopper and the optical signal
measured by synchronous detection. Both topographic signal from feedback control and
optical signal are recorded by  a computer, which controls the  XY scanning.

\begin{figure}[!htbp]
        \includegraphics[width=8.5cm]{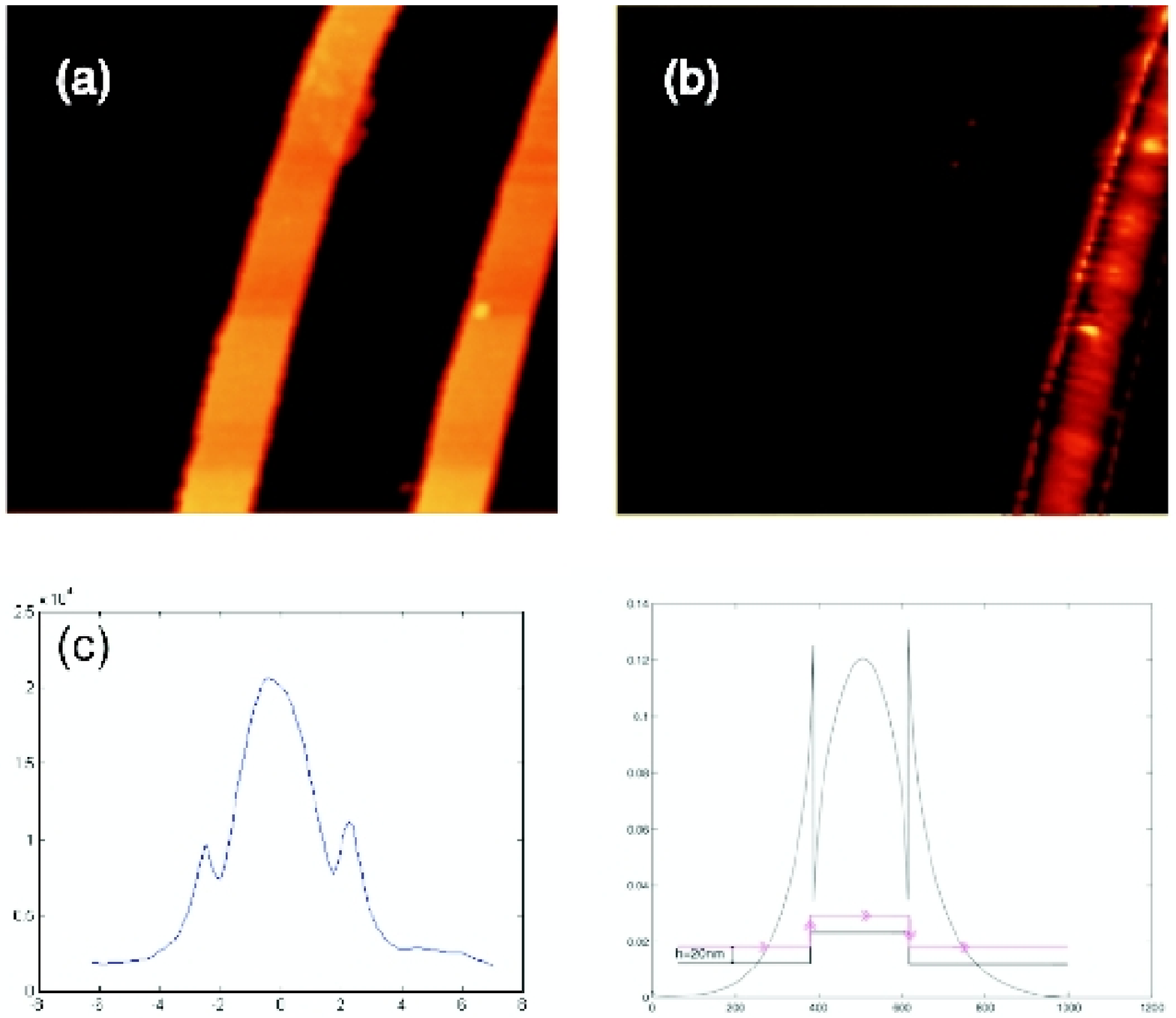}
        \caption{Near-field characterization of two bare waveguides. (a) Topographic
        image ($40 \times 40 \mu m^2$)
        (b) Corresponding optical image (c) Section of experimental optical image (d)
        Section of theorical
        optical image (BPM)}
        \label{fig.11}
\end{figure}

\subsubsection{Near-field characterization of bare waveguides}

Imaging the bare waveguides was a means to test the near-field microscope on a well
known intensity distribution and to control the guiding quality of the fabricated
waveguides. Fig.\ref{fig.11} (a) and (b) show respectively topographic and optical image
of two waveguides. Light has been injected only into the right waveguide. Optical image
shows that no light has been collected into the second waveguide despite the fact that
the two waveguides were $10 \mu m$ apart. The optical image (Fig.\ref{fig.11} (b)) and
its section (Fig.\ref{fig.11} (c)) show a minimum of intensity at the edges of
waveguide, which matches with the theorical optical field intensity section calculated
at constant distance by BPM calculation (Fig. \ref{fig.11} (e)).

\subsubsection{Near-field characterization of  a complete lattice}

\begin{figure}[!htbp]
        \includegraphics[width=9.0cm]{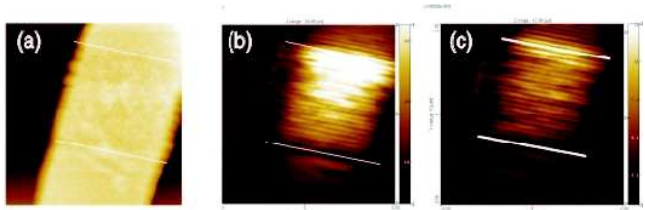}
        \caption{Near-field characterization of hole lattice on the waveguide
        at two wavelengths ((b) $\lambda =850nm$; (c) $\lambda =900nm$). (a)
        Topographic
        image ($5 \times 5 \mu m^2$). (b), (c) Corresponding optical images.}
        \label{fig.12}
\end{figure}

We have characterized at different wavelengths a lattice of $40 \times 40$ holes
(diameter: $d=200nm$, period: $a=360nm$) pierced on the waveguide. Even if transmission
spectra did not show any clear photonic band $LiNbO_3$, light penetration into such structures
have important variations with the wavelength. Two images (topographic and optical) at
two wavelengths (Fig.\ref{fig.12}), one in the theorical band gap ($\lambda =900nm$,
Fig.\ref{fig.12} (c)), the other out of the bandgap ($\lambda =850nm$, Fig.\ref{fig.12}
(b)) show two completely different behaviours. Topographic image (Fig.\ref{fig.12} (a)
and (c)) exhibits periodic nanostructures (between white lines). Light propagates from
top to bottom. Fig.\ref{fig.12} (b) shows a deep penetration of the light into the
structure. For the other wavelength, Fig.\ref{fig.12} (c) shows a bright zone  near the
upper limit of the hole lattice but light decreases rapidly inside the lattice. The
transmission efficiency (ratio output/input) of the lattice is five times greater at
$\lambda=850nm$ than at $\lambda=900nm$.

Fringes inside and in front of the lattice must be noticed: multiple reflections inside
such structures and standing waves outside could explain this phenomenon.

\begin{figure}[!htbp]
        \includegraphics[width=9.0cm]{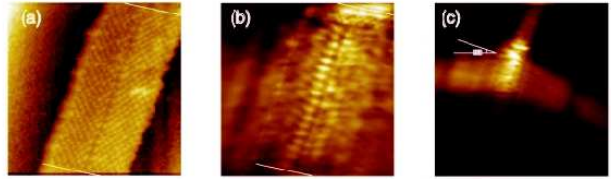}
        \caption{Near-field characterization of the hole lattice with a line defect on
        the waveguide
        at $\lambda =825nm$ (a) Topographic
        image ($10 \times 10 \mu m^2$). (b)Corresponding optical image.
        (c) Optical image ($40 \times 40 \mu m^2$) of the same structure. Measured angle
        of the diffracted beam $\theta =18.8^o$.}
        \label{fig.13}
\end{figure}

\subsubsection{Near-field characterization of a lattice with a line defect}

A second type of structure ($80 \times 40$ hole lattice with a line missing in its
center) has been characterized. It can be considered as a photonic waveguide, in which
the injection would require a taper to ensure an efficient connection to the
conventional waveguide, as its width is 10 times smaller.

As in the previous part, penetration and propagation of light into the structure depends
on the wavelength. Fig.\ref{fig.13} shows topographic and optical images at $\lambda
=825nm$ which is outside the theoretical bandgap. On the topographic image the missing
line is visible in the center of the waveguide. The optical signal exhibits a dark line
located on the missing holes and two lines of bright dots on both sides. This result
presents similarities with the numerical simulation Fig.\ref{fig.5} (b). These bright
lines are completed by two diffracted beam which are guided in the SiON layer on both
sides of the waveguide (Fig.\ref{fig.13}(c)).

\section{CONCLUSIONS}

Direct FIB milling has shown some limits (hole shape, etching depth) which have caused
imperfection in the fabricated photonic crystals. We are improving another similar way
of fabricating such nanostructures: FIB is used only to engrave a $200-250nm$ thick
metallic layer, which  is considered as a mask for RIE or deep-RIE. A shorter milling
time and a better aspect ratio are expected from this process.

In the present conditions the 2D approximation we used in FDTD simulation helped us to
chose the wavelength domain in the characterization but the imperfect structures had no
bandgap effect to be compared with the theoretical one. In spite of this difficulty the
characterization of imperfect photonic structures is an example of the diagnosis
delivered by near-field probing applied to complex 3D structures, the properties of
which are not predicted by a 2D model. Despite the imperfection of the fabricated
structures, the experimental and theoretical images of light propagation in a
line-defect lattice present interesting similarities.

\begin{acknowledgments}
Many thanks to \' Elo\"\i se Devaux (INIST, Strasbourg) for her technical assistance and to Maria-Pilar Bernal, Matthieu Roussey, Nad\` ege Bodin, Dominique Heinis for fruitful discussions.
\end{acknowledgments}

\end{document}